\documentclass{elsart}

\usepackage{graphicx}
\def\br{\begin{eqnarray}}
\def\er{\end{eqnarray}}
\def\be{\begin{equation}}
\def\ee{\end{equation}}
\def\a{\alpha}
\def\b{\beta}

\def\G{\Gamma}

\def\m{\mu}
\def\P{\Pi}

\def\d{\delta}
\def\<{\left\langle}
\def\>{\right\rangle}
\def\gc{\<{\frac{\alpha_s}{\pi}}G^{\mu\nu}G_{\mu\nu}\>}
\def\a2{\<A^{\mu}A_{\mu}\>}
\def\G{\Gamma}

\begin{document}

\begin{frontmatter}

\title{A dynamical gluon mass solution in Mandelstam's approximation}
\author[ift]{A. C. Aguilar}
\ead{aguilar@ift.unesp.br}
\author[ift]{A. A. Natale}
\ead{natale@ift.unesp.br}
\address[ift]{Instituto de F\'{\i}sica Te\'orica,
Universidade Estadual Paulista\\
Rua Pamplona 145,
01405-900, S\~ao Paulo, SP,
Brazil
}%
\date{\today}

\begin{abstract}

We discuss the pure gauge Schwinger-Dyson equation for the gluon propagator in the Landau gauge within an approximation
proposed by Mandelstam many years ago. We show that a dynamical gluon mass arises as a solution. 
This solution is obtained numerically in the full range of momenta that we have considered without 
the introduction of
any ansatz or asymptotic expression in the infrared region. The vertex function that we use follows a
prescription formulated by Cornwall to determine the existence of a dynamical gluon mass in the
 light cone gauge. The renormalization procedure differs from the one proposed by Mandelstam and allows for the possibility of a dynamical gluon mass.
Some of the properties of this solution, such as its dependence on $\Lambda_{QCD}$ and
its perturbative scaling behavior are also discussed.

\end{abstract}


\end{frontmatter}

\section{Introduction}

It is widely believed that Quantum Chromodynamics (QCD) is the theory which describes the strong
 interaction. For this theory we
know that  perturbation
theory has become a  reliable field theoretical method of calculating and predicting most of the
 quantities  in processes where high energies are
transferred between quarks and gluons.

This successful procedure to deal with the strongly interacting phenomena is known to be inadequate 
when it is applied to the infrared region.
There are phenomena at low energies such as dynamical chiral symmetry breaking, that, in
principle, could only be described when all orders of perturbation theory are taken into account,
 it means that they are necessarily
of a non-perturbative nature.

To bridge the gap between these two regions, infrared and ultraviolet, two main non-perturbative 
approaches are available, the lattice
theory which is based on discretization of space-time and a continuum one which makes use of an 
infinite tower of coupled integral
equations that contain all the information about the theory - the so called Schwinger-Dyson
 Equations (SDE).

In the continuum, we hope that the SDE provide an appropriate framework to study the transition 
from the perturbative to the non-perturbative
behavior of the QCD Green functions. However, its intricate structure only become tractable when
 we make some approximations.

Many attempts have been made to understand the gluon propagator behavior through SDE. In the 
late seventies Mandelstam initiated the
study of the gluon SDE in the Landau gauge \cite{mand}. Neglecting the ghost fields contribution 
and imposing cancellations of certain terms in the gluon polarization
tensor, he found a highly singular gluon propagator in the infrared. This enhanced gluon propagator
 was appraised for many years in the literature,
firstly because it provided a simple picture of quark confinement \cite{west}, since it is possible
 to derive from it an interquark potential that rises linearly
with the separation, and secondly because a gluon propagator, which is singular as $1/q^4$, has
 enough strength to support dynamical chiral
symmetry breaking, as it was claimed in the studies of the quark-SDE.  This approximation 
and its solution were extensively
studied by Pennington and collaborators \cite{Brown}.
However, these results are discarded by simulations of QCD on the
lattice at $95 \% $ confidence level \cite{mar}, where it is shown
that the gluon propagator is probably infrared finite.

Infrared finite solutions are also found in the Schwinger-Dyson approach, as result of different
 procedures. Many years ago, making use
of the ``pinch technique", Cornwall built up a gluon equation trading the conventional 
gauge-dependent SDE by one formed by gauge-independent
blocks. Analyzing this new equation, he obtained a gluon propagator endowed with a dynamical 
mass \cite{cornwall}. 

Recently, a quite extensive work on pure gauge SDE has been done
by the authors of Ref.\cite{alkofer} where they have shown that
when the ghost fields are taken into account, the  gluon propagator is suppressed and the ghost 
propagator is enhanced in the infrared region. Such solution was shown to satisfy the Kugo-Ojima 
confinement criterion \cite{kugo,alkugo}. These propagators exhibit an
infrared asymptotic power law behavior which is characterized by
a critical exponent $\kappa$. Axiomatic considerations \cite{kondo} and the 
latest results of the coupled gluon-ghost SDE seems to suggest
that $\kappa=0.5$ is allowed \cite{bl2}, signaling the possibility of
dynamical mass generation for the gluon.

All these solutions appear because different approximations were used and furthermore it is
 also perfectly possible that in the
same approximation more than one solution arise. It is interesting to note that, according to 
Mandelstam's work (see the comments
after Eq.(2.16) of Ref.\cite{mand}), a massive gluon solution was discarded from the beginning 
in his study.

It is important to stress at this point that a dynamical gluon mass does not break gauge invariance
 and is consistent
with {\sl massive} Slavnov-Taylor identities \cite{cornwall}. We would like to emphasize that
the presence of a dynamically generated
mass also does not mean that gluons can be considered as massive asymptotic states similar to 
dynamically generated quark mass does not
mean that quarks can be observed as massive asymptotic states. Why quarks and gluons are not 
observed as free states is the well known
problem of confinement. In the case of a theory with massive gluons we know that such theories 
admit  a vortex solution that may give a clue about the confinement
mechanism  \cite{cornwall,vortex}. Furthermore, a dynamically generated gluon mass is possibly
 connected to the existence of a QCD infrared fixed point \cite{ans}, whose presence has many
 phenomenological implications as nicely reviewed in Ref.\cite{brodsky}.

Our aim here is to revisit the gluon SDE within the Mandelstam approximation, in order to 
obtain a massive gluon solution. In section II, we
start building up  the gluon SDE, which embodies not just the full gluon propagator but also 
involves the  full triple gluon vertex. In order to allow
that a dynamical gluon mass takes place without breaking the relationship between  the Green's 
functions of different orders
which are imposed by the gauge invariance, the full triple gluon vertex behavior is modeled by a 
suitable Slavnov-Taylor identity.
Having found the gluon-SDE, the ultraviolet divergences will be removed by a subtractive renormalization 
procedure, in a similar way as
performed by Cornwall in Ref.\cite{cornwall}(although the equation is different), which introduces an arbitrary scale $\mu^2$ that
 can be related to the usual QCD
scale $\Lambda_{QCD}$. This discussion will appear in section III. We then solve the gluon equation
 by an iterative numerical procedure on the whole
range of momenta and present our results in section IV. In the present work we do not need to make
 any ansatz for the infrared solution.
Section V contains a discussion about the vacuum energy and stability of the solution. We draw our
 conclusions in section VI.

\vspace{0.5cm}

\section{The gluon equation in the Mandelstam's approximation}

\begin{figure}[ht]
\vspace{-1.0 cm}
\begin{center}
\includegraphics[scale=0.6]{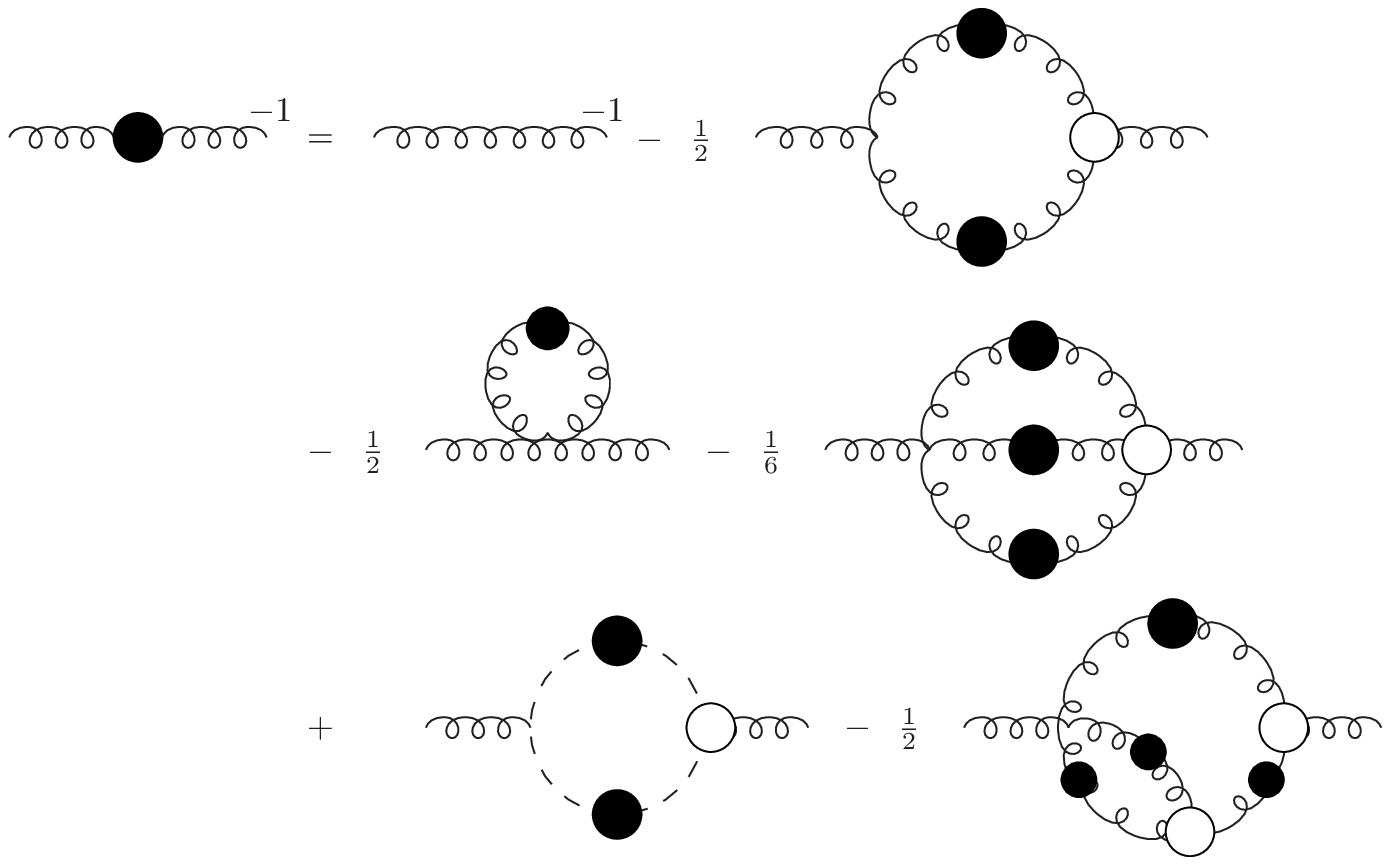}
\end{center}
\vspace{-8.5cm}
\caption{The complete Schwinger-Dyson equation for the gluon propagator without quarks. 
The spiral lines represent gluon
field and the dashed lines ghosts. The black blobs are the full propagators while the white 
ones are the full vertices.}
\label{fullesd}
\end{figure}

The SDE are coupled integral equations which relate all the Green's functions of the theory.  
To illustrate how intricate is its structure we can look at what are the Green functions which 
are involved in the full gluon  equation.  Neglecting the fermionic interactions we can
see, in Fig.(\ref{fullesd}), that the  gluon propagator is written in terms of itself, the 
full 3 and 4-point gluon vertex, $\Gamma_{\mu\nu\rho}$ and $\Gamma_{\mu\nu\rho\sigma}$, and also 
the full ghost propagator and the gluon-ghost coupling.

%%%%%%%%%%%%%%%%%%%%%%%% FIG. 1 %%%%%%%%%%%%%%%%%%%%%%%%%%%%%%%%%%%%%%
%%%%%%%%%%%%%%%%%%%%%%%%%%%%%%%%%%%%%%%%%%%%%%%%%%%%%%%%%%%%%%%%%%%%%%%

Actually, these unknown three and four point-functions obey their own SDE, which involve 
higher n-point functions which naturally must satisfy, in their turn, their own SDE. In fact,
 it is this entanglement of equations which makes  unavoidable the use of some truncation schemes.

 One famous truncation scheme is the Mandelstam's approximation where the fermion fields are 
neglected, since we believe that a pure Yang-Mills theory  must carry all the main features of QCD. 
Furthermore, based on perturbative results, the ghost fields are also neglected.
The justification, for the latter approximation is that the contribution which comes from 
the ghost field is supposed to be small even in the non-perturbative region. Therefore, this 
approximation can be represented pictorially by the graphics which compound  the first line of
the Fig.(\ref{fullesd}) and is written as \cite{mand}
\vspace{0.5cm}
\begin{eqnarray}
&&D^{-1\mu\nu}(k^2)=D_{0}^{-1\mu\nu}(k^2)  \nonumber \\
&&\hspace{1cm} +
g_{0}^{2}C_{2}\frac{1}{2}\int
\frac{d^{4}q}{(2\pi)^{4}}\Gamma_{0}^{\mu\rho\alpha}(k,-p,q)
D_{\alpha\beta}(q^2)D_{\rho\sigma}(p^2)\Gamma^{\beta\sigma\nu}
(-q,p,-k), \label{esd}
\end{eqnarray}
where $ p=k+q $, $ D_{0}$  and $ \Gamma_{0}$ are respectively  the propagator and
three-gluon vertex at tree level,
while $D$ and $ \Gamma $  are the two and three points full Green functions. In the case 
of the full gluon propagator in
Landau gauge, $D^{\mu\nu}$, we can write it as
\begin{equation}\label{prop1}
D^{\mu\nu}(q^2) = \left( \delta^{\mu\nu}
-\frac{q^{\mu}q^{\nu}}{q^2}  \right)\frac{{\mathcal Z}(q^2)}{q^2},
\end{equation}
where it will be useful to define the function, $D(q^2)$, in terms of the gluon
 renormalization function, ${\mathcal Z}(q^2)$.
\begin{equation}
D(q^2)=\frac{{\mathcal Z}(q^2)}{q^2}.
\label{func_d}
\end{equation}

Neglecting all contributions coming from ghosts fields, the Slavnov-Taylor identity 
between the three gluon vertex and the
inverse of the gluon propagator can be expressed in the following simple form
\begin{equation}
\label{slavnov}
k_{\mu}\Gamma^{\mu\nu\rho}(k,p,q)=  \frac{q^2}{{\mathcal{Z}}(q^2)}\left(\delta^{\nu\rho}
-\frac{q^{\nu}q^{\rho}}{q^2}\right) -
\frac{p^2}{{\mathcal{Z}}(p^2)}\left(\delta^{\nu\rho} -
\frac{p^{\nu}p^{\rho}}{p^2}\right).
\end{equation}

In order to allow that a massive gluon propagator will be also compatible with the 
Slavnov-Taylor identity expressed above, and supposing that  the
gluon renormalization function, ${\mathcal Z}(q^2)$ admits an expression of the following form
\begin{equation}
{\mathcal Z}(q^2)=\frac{q^2}{q^2+m^2},
\end{equation}
we note that it must be added new terms, that have massless poles, to the structure of
 the three gluon vertex, which, apart from a group theoretical factor, lead us to
a first modification that we should introduce in the construction of the full vertex, 
which is the one prescribed by Cornwall many years ago \cite{cornwall}
\begin{equation}\label{vpert}
\Gamma^{(m)}_{0\, \mu\nu\rho}(k,p,q)= (k-p)_{\rho}\delta_{\mu\nu}  +
 \frac{m^2}{2}\frac{k_{\mu}p_{\nu}
(k-p)_{\rho}}{k^2p^2} + \mbox{c.p.},
\end{equation}
where $c.p.$ means cyclic permutation, and, as discussed in Ref. \cite{cornwall},
 $\Gamma^{(m)}_{0\, \mu\nu\rho}$ is the vertex
for the massive theory.

As remarked by Corwnall \cite{cornwall}, during the procedure of construction of a vertex 
function which
automatically satisfied the Slavnov-Taylor identities, Ball and Chiu \cite{ball} have done a crucial assumption in order to
get a unique form for the longitudinal vertex. They supposed that the vertex should be free of kinematic singularities, since most of these singularities violate the general analyticity requirements of the vertex.  Therefore, 
tensorial structures  which have massless poles in three gluon vertex were explicitly excluded in their construction, despite the 
fact that they have already mentioned in the same work, that there are certain types which might naturally occur without the 
breakdown of analyticity in the
three gluon vertex, and one structure of this type is the one given by the second term 
in the right hand side of Eq.(\ref{vpert}).

It is important to bear in mind that the mass $m$ in Eq.(\ref{vpert}) has a momentum dependence 
and do not destroy the unitary behavior of the theory
\cite{cornwall}. Moreover, as stressed before, the role of the latter term from Eq.(\ref{vpert}), 
is  only to allow that the gluon propagator could
assume a non-zero value, {\it i.e. $D^{-1}(q^2=0)\neq 0$}, at the deep infrared region, without 
breaking the gauge invariance
imposed by the Slavnov-Taylor identity.  Exactly the massless poles of Eq.(\ref{vpert}) lead to
 the possibility of a mass gap, contrarily
to the solutions for the infrared gluon propagator behaving as $1/q^4$ found
in Ref.\cite{mand,baker} where a different vertex choice is made.

The next step in the Mandelstam approximation is to define the final expression for the full
 three gluon vertex. The form of the full three gluon vertex is \cite{Brown}
\begin{equation}
\Gamma^{\mu\nu\rho}(k,p,q) = \frac{1}{{\mathcal Z}(p^2)}\Gamma^{{(m)} \, \mu\nu\rho}_{0}(k,p,q).
\label{full}
\end{equation}

The use of the above full vertex, which is a combination of the Cornwall's and Mandelstam's
 prescriptions,
simplify even more the structure of gluon SDE than the use of the bare vertex, once it implies a 
cancellation
between  the gluon renormalization function which comes from a full gluon propagator and the one 
that comes from the full triple vertex.
\vspace{-2cm}
%%%%%%%%%%%%%%%%%%%%%%%% FIG. 2 %%%%%%%%%%%%%%%%%%%%%%%%%%%%%%%%%%%%%%
\begin{figure}[ht]
\begin{center}
\hspace{-2.0 cm}
\includegraphics[scale=0.6]{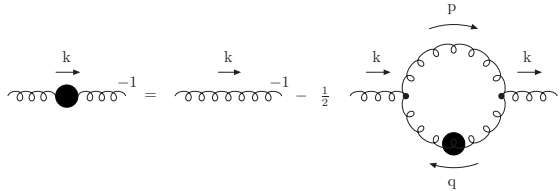}
\end{center}
\vspace{-12cm}
\caption{The gluon Schwinger-Dyson equation in the Mandelstam's approximation.}
\label{apmad}
\end{figure}
%%%%%%%%%%%%%%%%%%%%%%%%%%%%%%%%%%%%%%%%%%%%%%%%%%%%%%%%%%%%%%%%%%%%%%%
%

With the above expressions for the gluon propagators and vertices  and contracting this 
result with the projector, proposed by Brown and Pennington \cite{Brown},
\begin{equation}\label{proj}
  {\mathcal{R}}^{\mu\nu}(k) =\delta^{\mu\nu}
  -4\frac{k^{\mu}k^{\nu}}{k^2},
\end{equation}
the follow equation comes out,

\begin{eqnarray}\label{pennington}
\frac{1}{{\mathcal Z}(k^2)}= 1 &&+ \frac{g_{0}^2}{16\pi^2}
\int^{k^2}_{0}\frac{dq^2}{k^2} \left( \frac{7}{2}\frac{q^4}{k^4}
- \frac{17}{2}\frac{q^2}{k^2}
-\frac{9}{8}\right){\mathcal Z}(q^2) \nonumber \\&& \hspace{1.5cm}
+\frac{g_{0}^2}{16\pi^2}\int^{{\Lambda}^2}_{k^2}\frac{dq^2}{k^2}\,
\left( \frac{7}{8}\frac{k^4}{q^4} -
7\frac{k^2}{q^2}\right){\mathcal Z}(q^2),
\label{fesd}
\end{eqnarray}
where $g_0$ is the bare coupling, we use the color factor, $C_{2}=3$ and $\Lambda$ is an ultraviolet 
cutoff which was introduced in
order to render the integral finite.
In fact, the great advantage of the Mandelstam approximation is that all the angular integrals can be performed analytically
leading us to a much simpler equation.

A remark concerning the  massive term of the three gluon vertex is in order.  This equation is
 exactly the same one obtained in Ref.\cite{Brown}.
Although the three gluon vertex has an extra term, no contributions come from it. Therefore, as mentioned before, its only role is to
allow that the inverse  gluon propagator can be different from zero at the origin consistently with the Slavnov-Taylor identity.

\section{Renormalization}
\label{rm}

The procedure to perform the renormalization of SDE in the Mandelstam's approximation 
consists in introducing the  gluon and
vertex  renormalization constants, $Z_3$ and $Z_{g}$, respectively which will absorb ultraviolet divergences of the equation. Through
these constants we can define the  following renormalized quantities

\begin{eqnarray}
D(q^2,\Lambda^2)&=&Z_3(\mu^2,\Lambda^2)D_R(q^2,\mu^2)    \nonumber \\
g_0(\Lambda^2)&=&Z_g(\mu^2,\Lambda^2)g(\mu^2),
\label{renor}
\end{eqnarray}
where $D_{R}$ and $g$ are the renormalized gluon propagator and the renormalized coupling.

Substituting into  the SDE, Eq.(\ref{fesd}), the nonrenormalized quantities, $D$ and $g_0$, by the renormalized ones lead us to
\begin{equation}\label{renorm}
D_{R}(k^2)= \left[k^2Z_{3} +
Z_{3}^{2}Z_{g}^{2}\frac{g^2(\mu^2)}{16\pi^2}{\mathcal {
I}}_{D_{R}}(k^2)\right]^{-1},
\end{equation}
where ${\mathcal{I}}_{D_R}(k^2)$ is given by
\begin{eqnarray}\label{funcional}
{\mathcal {
I}}_{D_{R}}(k^2)&=&\int^{k^2}_{0} dq^2\left(
\frac{7}{2}\frac{q^4}{k^4} - \frac{17}{2}\frac{q^2}{k^2}
-\frac{9}{8}\right)q^2D_{R}(q^2) \nonumber \\
&&\hspace{3cm}+\int^{{\Lambda}^2}_{k^2} dq^2\, \left(
\frac{7}{8}\frac{k^4}{q^4} - 7\frac{k^2}{q^2} \right)q^2D_{R}(q^2).
\end{eqnarray}

The vertex renormalization constant $Z_g$ can be eliminated from the Eq.(\ref{renorm}), using
 the identity, $Z_{g}Z_{3}=1$, which is  only valid in this approximation \cite{alk}. However, there
 still remains the gluon renormalization
constant $Z_3$ that is a naturally divergent quantity.

Despite all efforts to obtain a totally consistent renormalization of the gluonic SDE, be it in this
 truncation or even beyond the
Mandelstam approximation, we know that the renormalization of SDE is highly nontrivial. 
To illustrate this claim we notice that only in the
nineties, Curtis and Pennington pointed out a truncation scheme, for QED, which is gauge 
independent and also respect the multiplicative
renormalizability \cite{CP}. In the following we consider $Z_3$ as a factor which
renders the divergent terms of the gluon SDE in a finite one through a subtractive renormalization.

After imposing $Z_{g}Z_{3}=1$ in Eq.(\ref{renorm}) we are left with the following
equation
\begin{eqnarray}\label{presc}
D_{R}(k^2)^{-1}&=& k^2\left[ Z_3 + \frac{\alpha (\mu^2)}{4\pi} \int^{{\Lambda}^2}_{k^2} dq^2\, 
\left(- 7\right)D_{R}(q^2) \right] \nonumber \\
&+& \frac{\alpha (\mu^2)}{4\pi}
\int^{k^2}_{0} dq^2\left(
\frac{7}{2}\frac{q^4}{k^4} - \frac{17}{2}\frac{q^2}{k^2}
-\frac{9}{8}\right)q^2D_{R}(q^2)  \nonumber \\
&+& \frac{\alpha (\mu^2)}{4\pi}\int^{{\Lambda}^2}_{k^2} dq^2\, 
\frac{7}{8}\frac{k^4}{q^2}D_{R}(q^2)
\nonumber \\
\end{eqnarray}
where $ \alpha(\mu^2)=\frac{g^2(\mu^2)}{4\pi}$. $Z_3$ behaves as $1$ plus an infinite
piece. The factor $1$ is consistent with the perturbative behavior, and the infinite part of $Z_3$ must cancel the infinite part that comes out from the first integral in Eq.(\ref{presc}), which
is the only divergent integral in the above equation. 

It is important to notice that Eq.(\ref{presc}) is not of the same form as the one
shown in Ref.\cite{cornwall}, as well as we shall not follow the same procedure adopted up to now in the many papers to
renormalize the Mandelstam equation. In the previous papers where this equation
was solved it was verified that the dressing of the gluon propagator could behave as 
\begin{equation}
Z(k^2)= \frac{1}{A + B k^2 + J(k^2)}
\label{ABJ}
\end{equation}
The numerical solutions were then obtained assuming $A=0$ and $B=1$, which are not independent conditions. Note that these conditions are necessary to obtain a
solution compatible with the Slavnov-Taylor (ST) identities in the case of a massless
gluon. As discussed in the previous section, it is possible to introduce a new piece in the 
gluon vertex that makes the ST identities compatible with dynamical mass
generation for the gluons, alleviating any condition on $A$ and $B$.   

We will face $Z_3$ as the term that will eliminate the
infinite contribution of the following expression
\begin{equation}
Z_{3} - \frac{\alpha(\mu^2)}{4\pi}\int^{\Lambda^2}_{k^2} \, dq^2 
\,7 D_{R} (q^2) \,\,\, .
\label{refin}
\end{equation}
Of course, in the above subtraction there remains a finite contribution.
$D_R(q^2)$ must be known to perform the above calculation, but it is going to be
known only after we impose the renormalization conditions

\begin{equation}
{\mathcal{Z}_{R}}(\mu^2)=1,
\end{equation}
or equivalently
\begin{equation}
D_{R}^{-1}(\mu^2)=\mu^2.
\label{cond}
\end{equation}

The role of the above equation is to guarantee that exactly at the point $\mu^2$ we 
recover the bare
perturbative behavior, $D_R(q^2) = 1/q^2$. For this reason it is important that
the value of $\mu^2$ be fixed at a typical
perturbative scale, {\it i.e.}
it must satisfy the condition  $\mu^2 \gg \Lambda_{QCD}$.

Let us now discuss the prescription for the finite contribution to Eq.(\ref{refin}).
The solution usually found for the Mandelstam gluon propagator is of the form
$D_1(k^2)\approx 1/k^4$. But in principle, as long as we do not impose $A=0$ and $B=1$ as discussed
after Eq.(\ref{ABJ}), we cannot discard a solution behaving like $D_2(k^2) \approx 1/[(k^2+m^2)\ln (k^2+m^2)]$, which is a massive solution similar to the one found by Cornwall in Ref.\cite{cornwall}.  

If $D_R(q^2)$ behaves as $D_1(q^2)$ it is natural to have a finite contribution like
$\kappa_1/k^2$ in the integration of Eq.(\ref{refin}). The same is true for high
values of $k^2$ if $D_R(q^2)$ behaves as $D_2(q^2)$, since the integral
that appears in Eq.(\ref{refin}) will generate terms like $\ln\ln \Lambda^2$,
 $\ln\ln k^2$ and again a term proportional to $m^2/k^2$. It is also not likely that
the integral in Eq.(\ref{refin}) will give finite contributions like $\delta_n/(k^2)^n$ with $n\geq 2$, because this behavior is not consistent with any expected form of $D_R (q^2)$. As $\kappa_1$ and $m^2$ have exactly the same dimension, the simplest form of Eq.(\ref{refin})which is compatible with both behaviors 
is
\begin{equation}
Z_{3} - \frac{\alpha(\mu^2)}{4\pi}\int^{\Lambda^2}_{k^2} \, dq^2 
\, 7 D_{R} (q^2) = 1 + \frac{\kappa}{k^2} \,\, ,
\label{re2}
\end{equation}
where $\kappa$ has squared mass dimension and is going to be fixed by the 
renormalization procedure ($D_{R}^{-1}(\mu^2)=\mu^2$).

A more complex expression on the right-hand side would introduce new constants
besides $\kappa$, which could be fixed only in an {\sl ad hoc} fashion.
The $1$ in the right-hand side of the above expression corresponds to the first $Z_3$ term.
The choice we made in Eq.(\ref{re2}) does not obey
the conditions $A=0$ and $B=1$ that we discussed before, and, in  
principle, it allows even for a massive gluon solution.
This possibility is an actual one as long as we obtain
a stable solution for the final equation. 

In all the procedures used up to now to
solve the Mandelstam equation, it has been assumed a cancellation of
certain terms, and there is not any discussion about the finite terms that
survive renormalization. It is known that in perturbative calculations such finite
terms are nothing else than small quantum corrections which modifies the tree level value 
of the renormalized quantity. On the other hand the physical quantities which are
related to the non-perturbative dynamical mass generation will be exclusively generated by the quantum corrections. This means that such finite terms surviving renormalization are fundamental to the final result. 

With  the definition described by the Eq.(\ref{re2}) and imposing the renormalization condition
(\ref{cond}) into Eq.(\ref{presc}) we
can obtain the following expression for the parameter $\kappa$,
\begin{eqnarray}
\kappa= -&& \frac{\alpha(\mu^2)}{4\pi} \int^{\mu^2}_{0}dq^2 \left( \frac{7q^4}{2\mu^4}-
\frac{17q^2}{2\mu^2}-\frac{9}{8} \right)q^2D_R(q^2) \nonumber \\
&& \hspace{3cm}- \frac{\alpha(\mu^2)}{4\pi} \int^{\Lambda^2}_{\mu^2} dq^2 \frac{(7\mu^4)}{8q^4}q^2D_R(q^2)
\label{kexp}
\end{eqnarray}
and, consequently, the final expression for the inverse of the gluon propagator can be written as
\begin{eqnarray}
D_R^{-1}(k^2)=\kappa + k^2 +&&  \frac{\alpha(\mu^2)}{4\pi}\int^{k^2}_{0}dq^2 
\left( \frac{7q^4}{2k^4}-\frac{17q^2}{2k^2}-\frac{9}{8} \right)q^2D_R(q^2) \nonumber \\
+&& 
\frac{\alpha(\mu^2)}{4\pi} \int^{\Lambda^2}_{k^2} dq^2 \frac{7k^4}{8q^4}q^2D_R(q^2).
\label{sde_final}
\end{eqnarray}

Since we are now dealing only with renormalized quantities, in the sequence we will dismiss the subscript $R$ in the Green functions in order to get a more compact notation.
It is interesting to note that the renormalization constant $Z_{3}$ as it is established
 by Eq.(\ref{re2}) is proportional 
to $1$ plus a function of $\mu^2$ and $\Lambda^2 $.
Such behavior is compatible with the expected weak coupling expansion for this constant.

As already stated, the renormalization procedure in Schwinger-Dyson equations  has an intricate 
structure and other
choices have already been applied in these equations, such as the ``plus prescription'' 
\cite{pb} where the contributions
which violate the massless Slavnov-Taylor identity would be subtracted out of the right hand 
side of Eq.(\ref{fesd}) or even in
more elaborated cases, as the gluon-ghost coupled system, where we have to deal with a bigger
 number of renormalization
constants, different approaches can be considered \cite{alkofer,bloch}. As long as we do not
 have an exact procedure for the renormalization
problem in non-Abelian SDE, we have to face this prescription as one more try, that certainly
can be improved,  where  the quantitative perturbative
behavior of the renormalization constant $Z_3$ is reproduced.

Our prescription in Eq.(\ref{re2}) acts as a seed for the dynamical gluon mass generation,
since $\kappa$ defined by Eq.(\ref{kexp}) is precisely a constant, like the $A$ term in
Eq.(\ref{ABJ}). Of course, this is not the most general prescription, and it does not
reproduce the exact perturbative ultraviolet behavior of the gluon propagator. This behavior
can be obtained if we use the following prescription
\begin{equation}
Z_{3} - \frac{\alpha(\mu^2)}{4\pi}\int^{\Lambda^2}_{k^2} \, dq^2 
\, 7 D_{R} (q^2) = 1 + \frac{\kappa_1}{k^2}+ \frac{4\pi}{\beta_0}\ln{\frac{k^2}{\kappa_2}} \,\, .
\label{re22}
\end{equation}
It is clear that the last term of Eq.(\ref{re22}) arises naturally from the integration
of the perturbative propagator ($D(q^2)=1/q^2$) in Eq.(\ref{refin}). However,   
the new constant ($\kappa_2$) cannot be fixed, since we have only one renormalization 
condition. In the next section we shall discuss the effect of a term like this one.

Before starting the numerical calculation of the gluon propagator, it is useful to remind the 
definition of the running coupling that will be valid from the non-perturbative  to the perturbative region. 
Remembering that, in the Mandelstam
approximation, $Z_3Z_g =1 $, it follows from Eq.(\ref{renor}) that
\begin{equation}
\alpha(q^2) = \alpha(\mu^2)\left[\frac{{\mathcal Z}(q^2)}{{\mathcal Z}(\mu^2)}\right]^2,
\label{npr}
\end{equation}
where we assume that ${\mathcal Z}(\mu^2)=1$.

It is important to keep in mind that, in QCD, we have the possibility to define the running 
coupling constant using the different vertices
of the theory, such as, the three gluon, four gluon, ghost-gluon or even the fermion-gluon vertices. Despite
the fact that in the perturbative QCD all these constructions, based on  different vertices,
 converge to the same behavior for the
running coupling, since the Slavnov-Taylor identities are preserved, the same does not happen in 
the low energy scale, due
to the complex renormalization procedure in the infrared region of QCD, which cause the loss of  the multiplicative renormalizability.  As
a consequence we may not have a unique definition for the running coupling in the infrared region. For this reason it is important  to stress that, in this approximation, our QCD running coupling  is constructed on the basis of three-gluon vertex, since  all the others vertices do not appear in
this approximation.

In the sequence we shall need the perturbative expression for the running coupling,
which is given by 
\begin{equation}
\alpha(q^2)= \frac{4\pi}{\beta_0\ln\left(\frac{q^2}{\Lambda_{QCD}^2} \right)},
\label{alpha_pert}
\end{equation}
where $\Lambda_{QCD}$ is the usual scale for QCD and $\beta_0$ is the first coefficient of the 
the Callan-Symanzik $\beta$ function,  $\beta(g)=\mu(dg/d\mu)$, which is written as
\begin{equation}
\beta(g) = - \beta_0 \frac{g^3}{16\pi^2} -      \beta_1 \frac{g^5}{(16\pi^2)^2} + \ldots,
\label{beta}
\end{equation}
where the $\beta_0$ and $\beta_1$ coefficients are given by
\begin{eqnarray}
\beta_0 &=& 11- \frac{2}{3}n_f \nonumber \\
\beta_1 &=& 102- \frac{38}{3}n_f.
\label{coef}
\end{eqnarray}

Since we neglected the fermions fields, we have that $n_f = 0$ which lead us to  $\beta_0= 11$ at one loop.

\section{The massive solution}

It has been shown in several works that it is not an easy task to obtain an analytical solution 
for the Eq.(\ref{sde_final}) \cite{mand,Brown,alkofer}
and therefore the only  possibility to face this problem is through a numerical
approach.

Our aim here is to find consistent solutions over the whole momentum range without impose,
 neither in the infrared region nor in
the ultraviolet, any previous asymptotic behavior obtained from an expansion of the gluon
 renormalization function, ${\mathcal Z}(q^2)$,
at small $q^2$ or obtained from the known perturbative behavior.

For this reason, we apply for the integral equation, given by Eq.(\ref{sde_final}), an iterative 
numerical method, starting with a trial function which can be too remote from the exact solution.

To implement so, it is convenient to introduce the following variables $x = k^2$ and $y =q^2$ in 
the Eq.(\ref{sde_final}), and we
also recall that since the beginning of this article all variables are in the Euclidean space, with these changes we
can write Eq.(\ref{sde_final}) as
\begin{equation}
\hspace{-0.5cm}D^{-1}(x)= \kappa + x +  \lambda\int^{x}_{0}dy \left( \frac{7y^2}{2x^2}-\frac{17y}{2x}-\frac{9}{8}
 \right)yD(y) + \lambda \int^{\Lambda^2}_{x}dy \left(\frac{7x^2}{8y^2}\right)yD(y) 
\end{equation}
where
\begin{equation}
\kappa = - \lambda \int^{\mu^2}_{0}dy \left( \frac{7y^2}{2\mu^4}-\frac{17y}{2\mu^2}-\frac{9}{8} 
\right)yD(y) - \lambda \int_{\mu^2}^{\Lambda^2}dy \left(\frac{7\mu^4}{8y^2}\right)yD(y),
\label{sde_final1}
\end{equation}
with $\lambda = \alpha(\mu^2)/{4\pi}$.

In order to study, in more details, the small $x$ region, we use a logarithmic grid for the 
variables $x$ and $y$, which allow us to vary the momentum from the deep infrared to the ultraviolet
 region.  Such logarithmic grid split the whole momenta range in two regions: the infrared region - 
defined by the range $[0, \mu^2]$ and the ultraviolet that comprehends the range $[\mu^2,\Lambda^2]$.
 The aim of this separation is to allow us to set  ${\mathcal Z}(\mu^2)=1$ or equivalently, 
$D^{-1}(\mu^2)=\mu^2$ \cite{krein}.

We start with a trial function $D(x)$ and use a cubic spline interpolation for generating 
the values of $D(y)$ which will be utilized in the right hand side of the Eq.(\ref{sde_final1})
 and in the sequence we compute this integral through  the Adaptive Richardson-Romberg extrapolation.

The initial trial function is compared with the result which was obtained after the integration
 and the convergence criteria to stop the numerical code is to impose that the difference 
 between input and output functions must be smaller than $10^{-4}$.

It is important to stress that we have tested variations of initial guesses and verified that 
our results are completely independent of the trial function imposed for $D(x)$.

We also analyzed  the solution proposed in Ref.\cite{Brown} and we noticed that it can 
only be reproduced if we consider exactly the same momentum range showed in their Fig.3
 (as is already stressed in Ref.\cite{alk})  and use a trial function which is very close
 to the result found in \cite{Brown}. Such exigency reflects the instability of this solution, 
because if we extend the numerical range or even start from a different guess we do not recover 
the divergent $1/k^4$ behavior.

Our input data are the renormalization point, $\mu^2$, and the coupling constant defined at 
this point, $\alpha(\mu^2)$, which has the effect of fixing the value of the QCD scale,
 $\Lambda_{QCD}$, through the Eq.(\ref{alpha_pert}).

As stressed in the Sec.(\ref{rm}), the value of $\mu^2$ must be fixed at a typical 
perturbative scale in order to recover the high energy behavior of the gluon propagator. With the 
aim of analyze the dependency of our solution on the renormalization point, we vary the values  
of $\mu^2$ and $\alpha(\mu^2)$ within the range $[10 \,\mbox{GeV}^2,30 \,\mbox{GeV}^2]$
 and $[0.20,0.25]$  respectively. Such variation correspond  to run the $\Lambda_{QCD}$ parameter 
from $182\,\mbox{MeV}$ to $557\,\mbox{MeV}$ as we show in the Table(\ref{t1}).

\begin{figure}[ht]
\begin{center}
\includegraphics[scale=0.8]{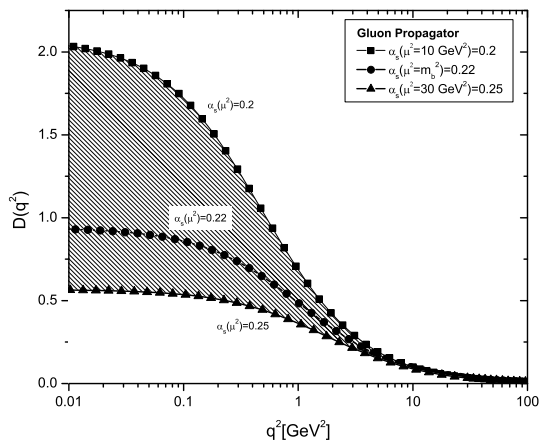}
\end{center}
\caption{Gluon propagator, $D(q^2)$, as function of momentum $q^2$ for different scales.  
The line + square curve was obtained when $\alpha(10 \, \mbox{GeV}^2)=0.2$ which corresponds 
 $\Lambda_{QCD}=  182\, \mbox{MeV}$, while in the line + triangle curve,
 $\alpha(30 \, \mbox{GeV}^2)=0.25$, which leads to $\Lambda_{QCD}=557\, \mbox{MeV} $. The
 shadowed area delimits the curves with $\Lambda_{QCD}$ varying within the range
 $ [182 \, \mbox{MeV}, 557 \, \mbox{MeV} ]$. The central curve (line + circle) was obtained 
when we fix the renormalization point, $\mu^2$, at bottom quark mass,
 $m_b^2=(4.5)^2 \, \mbox{GeV}^2$ with the central value of $\alpha(m_b^2)=0.22$. }
\label{f1}
\end{figure}

The curves produced by these different scales are plotted in the Fig.(\ref{f1}) where it is shown
 the gluon propagator, $D(q^2)$, as a function of the momentum $q^2$.
The external curves  delimit  the lower and the higher values of the $\Lambda_{QCD}$
 in the range  mentioned above. The other set of input, shown in the Table(\ref{t1}), reproduce 
the same qualitative behavior and they are restricted to the shadow band.

We have also computed the case where $\alpha(\mu^2)$ is fixed at the bottom quark
 mass, $m_b^2=(4.5)^2 \, \mbox{GeV}^2$, and its central value is  $\alpha(m_b^2) = 0.22 $ \cite{PDG}, 
such solution is represented by the curve ``line + circle" displayed on the Fig.(\ref{f1}).

%%%%%%%%%%%%%%%%%%%%%%%% FIG. 3 %%%%%%%%%%%%%%%%%%%%%%%%%%%%%%%%%%%%%%
%%%%%%%%%%%%%%%%%%%%%%%%%%%%%%%%%%%%%%%%%%%%%%%%%%%%%%%%%%%%%%%%%%%%%%%

We have run our numerical code efficiently within a momenta range
 of twelve orders of magnitude where the typical momenta values which were utilized vary 
 from $10^{-6}$ to $10^6 \, \mbox{GeV}^2$. We set the renormalization point at $m_b^2$,
 which is located approximately in the middle of our typical momenta range, and where
 the physical quantities can be certainly described by the perturbative theory. 

It is interesting to provide an analytic expression for the gluon propagator, 
in order to analyze the gluon mass values obtained in infrared region, when we set the 
renormalization point, $\mu^2$, at different values.  
With this aim we fit our numerical data by an Euclidean massive propagator 
expressed by \cite{cornwall}

\begin{equation}
D(q^2)= \frac{4\pi}{\beta_0\left(q^2 + m^2(q^2)\right)\ln\left(\frac{q^2 + 4m^2(q^2)}{\Lambda_{QCD}}\right)},
\label{corfit}
\end{equation}
where
\begin{equation}
m^2(q^2)= m^2_0\left[ \frac{\ln\left(\frac{q^2 + 4m_0^2}{\Lambda_{QCD}^2}\right)}
{\ln\left(\frac{4m_0^2}{\Lambda_{QCD}^2}\right)}\right]^{-12/11}.
\label{masscor}
\end{equation}

We have also considered the simpler expression

\begin{equation}
D(q^2)= \frac{1}{q^2 + {\mathcal M}^2(q^2)},
\label{prop}
\end{equation}
where the dynamical mass ${\mathcal M}^2(q^2)$ is described by
\begin{equation}
{\mathcal M}^2(q^2)= \frac{{m^{\prime}}^4}{q^2+ {m^{\prime}}^2}.
\label{ope}
\end{equation}

The last fit can be motivated by the gluon polarization tensor 
behavior at high energies, which can be predicted by OPE as~\cite{lav}
\be
\P_{OPE} (P^2) \sim - \frac{34 N \pi^2}{9(N^2-1)}
\frac{\gc}{P^2}.
\label{piuv}
\ee

On the other hand recently there has been a lot of discussion about a possible bilinear
 condensate of
the gluon field \cite{dudal}. Such condensate, $\a2$, would be responsible for a mass term
 appearing in the
infrared gluon polarization tensor \cite{dudal}.  Therefore the fit provided by Eq.(\ref{ope}) is 
just the simplest
way to account for the different condensate contributions to the polarization tensor, from where
 we could expect
approximate relations of the form $ {m^{\prime}}^2 \propto \a2$ or $ {m^{\prime}}^4 \propto \gc$. We will discuss 
such type of
relation in the next section.  

We can apply these simple fits for all curves shown in the Fig.(\ref{f1}) and in all cases  
it is remarkable the agreement found from the deep infrared region up to the ultraviolet 
regime, using a fit which has a unique
 parameter, $m_0$ or $m^{\prime}$.

\begin{figure}[ht]
\begin{center}
\includegraphics[scale=0.8]{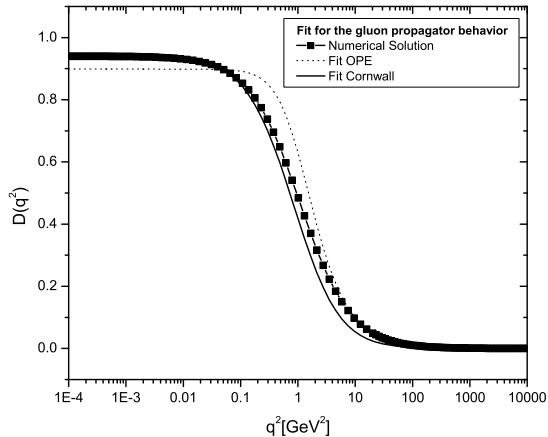}
\end{center}
\caption{Numerical solution for the gluon
propagator, $D(q^2)$, versus momentum $q^2$ for
$\alpha(m^2_{b})=0.22$. We compare this numerical solution
with the fit given by Eq.(\ref{prop}), where  ${m^{\prime}}^2= 1.11 \, \mbox{GeV}^2$, as well as the one given by the Eqs.(\ref{corfit}) and (\ref{masscor}) when  ${m_0}^2= 0.82 \, \mbox{GeV}^2$.}
\label{fig5}
\end{figure}

In particular, we plot in the Fig.(\ref{fig5}) the numerical solution for the gluon propagator, 
$D(q^2)$, when $\mu^2=m_b^2$  together with the curve obtained through our fit given by  
Eqs.(\ref{prop}) and (\ref{ope}) when  ${m^{\prime}}^2= 1.11 \, \mbox{GeV}^2$ as well as
the fit provided by the Eqs.(\ref{corfit}) and (\ref{masscor}) when  ${m_0}^2= 0.82 \, \mbox{GeV}^2$.

%%%%%%%%%%%%%%%%%%%%%%%% FIG. 4 %%%%%%%%%%%%%%%%%%%%%%%%%%%%%%%%%%%%%%
%%%%%%%%%%%%%%%%%%%%%%%%%%%%%%%%%%%%%%%%%%%%%%%%%%%%%%%%%%%%%%%%%%%

In particular we can easily extract the value of ${m^{\prime}}$ for the others curves displayed in the Fig.(\ref{f1}),
 to do so we basically have just to note that  in the limit of $q^2\rightarrow 0$,  
Eqs.(\ref{prop}) and (\ref{ope}) of our gluon propagator reduces to $D(q^2\rightarrow 0) = 1/{m^{\prime}}^2$, 
and therefore the inverse of ${m^{\prime}}^2$ is given by the value of the point in what the  gluon
 propagator curves  cross the  y axis in Fig.(\ref{f1}).

We verified that the value of $m_0$ and ${m^{\prime}}$ depend on the choice  of the
 renormalization point, $\mu^2$, however we must remember that when we change its value, actually
 what we are really changing is the scale of the theory, once that $\mu^2$ and $\Lambda_{QCD}$ are
 linked by Eq.(\ref{alpha_pert}). For this reason what matters is the analysis
 of the ratio $m_0/\Lambda_{QCD}$ or ${m^{\prime}}/\Lambda_{QCD}$ which, in principle, give to us a better idea  about
 the true dependency on the renormalization point of our solution.        

Fixing  the coupling constant and running the renormalization point we can see  from the 
Table(\ref{t1}) that the aforementioned ratios practically do not vary for the set of coupling constants shown in the table. This means that the ratios $m_0/\Lambda_{QCD}$ 
 ${m^{\prime}}/\Lambda_{QCD}$ are quite stable in our procedure. 

\begin{center}
\begin{table}[ht]
\caption{\label{t1} Values of the renormalization point, $\mu^2$, and coupling constant, 
$\alpha(\mu^2)$, used as input data in the Eq.(\ref{sde_final1}). In the third column, we have the values  of $\Lambda_{QCD}$  computed  with  the usual perturbative value of $\beta_0=11$. The values of the ratios
 $m_0/\Lambda_{QCD}$ and $m^{\prime}/\Lambda_{QCD}$ are also shown in the last two columns.} 

\begin{tabular}{ccccc}
   \hline
   $\alpha(\mu^2)$ & $\mu ^2$ & $\Lambda_{QCD}$& ${m_0}/{\Lambda_{QCD}}$& ${m^{\prime}}/{\Lambda_{QCD}}$ \\
            &        &  $(\beta=11)$   &  Cornwall & OPE  \\
  \hline
$0.20\qquad$&$10\,\mbox{GeV}^2$  & $182 \,\mbox{MeV}\qquad$ &$3.18$& $3.90$\\
$0.20\qquad$&$20\,\mbox{GeV}^2$  &$257 \,\mbox{MeV}\qquad$ &$3.18 $& $3.90$ \\
$0.20\qquad$&$30\,\mbox{GeV}^2$  &$315 \,\mbox{MeV}\qquad$ &$3.18$& $3.90$ \\
$0.22\qquad$&$m_b^2=20.25\, \mbox{GeV}^2$ & $335 \,\mbox{MeV}\qquad$& $2.70$ &$3.15$ \\
$0.25\qquad$&$10\,\mbox{GeV}^2$  &$321 \,\mbox{MeV}\qquad$ &$2.23 $& $2.45$ \\
$0.25\qquad$&$20\,\mbox{GeV}^2$  &$455 \,\mbox{MeV}\qquad$ &$2.28$& $2.44$ \\
$0.25\qquad$&$30\,\mbox{GeV}^2$  &$557 \,\mbox{MeV}\qquad$ &$2.23$& $2.43$ \\
\hline
\end{tabular}
\end{table}
\end{center}

We have also checked the effect of the prescription shown in  Eq.(\ref{re22}) attributing
random values to $\kappa_2$. We also obtained massive solutions with a better ultraviolet
behavior. The numerical curves are totally identical to the ones in Fig.(\ref{f1}) but 
they scale up or down by a constant factor as we change  the value of $\kappa2$. The only way to 
eliminate such
constant is forcing point by point of the numerical solution in the large momentum region to
match with the curve given by the known asymptotic perturbative solution. 

We now turn to the infrared behavior of the coupling constant which was built based on
 the triple gluon vertex as outlined above. By imposing the limit $q^2 \rightarrow 0$ in the
 Eq.(\ref{npr}) where the renormalization function, ${\mathcal Z}(q^2)$, can be extracted from 
 the fit which is expressed by Eq.(\ref{prop}) and (\ref{ope}) we clearly obtain a vanishing 
coupling in the deep infrared region within the  Mandelstam approximation. Although it may look 
surprising to  find that in a confining theory the IR coupling constant goes to zero, it is important to stress again that, for QCD, we have distinct definitions for $\alpha(q^2)$ which are based on different vertices of the theory,  moreover it is interesting to say that using the same vertex, a lattice QCD simulation found the same vanishing behavior for the coupling \cite{shirkov}. Of 
course, the introduction of ghosts can modify this result, and
it seems unlikely then that this limit  does reflect the true behavior of the coupling constant 
since we know that it must develop a non-trivial fixed point in QCD infrared region \cite{ans}.   

Following the same procedure that are perfomed here, the qualitative behavior of the gluon
propagator does not change when the ghosts 
fields are included when we study the coupled SDE for the gluon and ghost \cite{gg}. We believe 
that the main role of the ghost fields in covariant gauges is to guarantee that the coupling 
constant will be  finite and different from zero in the infrared regime, although it is quite
 hard to obtain numerically this freezing of the coupling without imposing any previous asymptotic 
form for the gluon and ghost propagators .

\section{Vacuum energy and stability of the solution}

In this section we would like to discuss some points about the
stability of the massive solution in this approximation. We call
attention to the fact that our solution is obtained numerically in
the full range of momenta. Other solutions for this kind of
equations in general assume one particular form of the solution in
the far infrared region and adjust free parameters  in a ``in" and
``out" procedure. We notice that our numerical code finds
stability for the $1/q^4$ solution found in Ref.\cite{Brown} only
when we enter in the code a seed basically given by the final
result of Ref.\cite{Brown}, otherwise there is no convergence up
to a large computing time. We credit this behavior to the fact the
$1/q^4$ must be an unstable solution of the SDE in this
approximation, because any seed that is slightly away from the
result of Ref.\cite{Brown} does not lead to a convergent
calculation. Unfortunately we cannot say more than that about the
stability of the $1/q^4$ solution.

It is possible to use some methods of integral equations to study the existence and
 stability of the solutions as
in Ref.\cite{atkinson}, but these methods may depend on the many approximations that are
 necessary to
make in order to obtain a tractable equation before they can be applied. We believe that the
 best to be done is
to compute the vacuum energy for composite operators \cite{cornja}. This vacuum energy is a
function of the full propagator and vertices of the theory~\cite{cornja,cornor}. The idea is 
simply to recall that
the vacuum energy will select the solution that leads to the deepest minimum of energy as 
discussed in Ref.\cite{mont}
in the case of pure gauge QCD. If we follow the results of Ref.\cite{mont} we can foresee that the massive solution
is the one selected by the vacuum. However we can do more than that and we will show that the computation of
the vacuum energy obtained in the previous section also leads to a consistent value for the gluon condensate.

We will briefly outline the calculation of the vacuum energy with the formalism of the effective potential for
composite operators \cite{cornja}.  It will also be computed according to the Mandelstam's 
approximation,
what is equivalent to neglect diagrams with fermions, ghosts and the quadrilinear gauge coupling.
 The details can
be obtained in Ref.\cite{gorbar}.   The effective potential with the approximations already
 discussed
has the form~\cite{cornja}
\begin{equation}
V(D) =  \frac{\imath}{2} \int \frac{d^4p}{(2\pi)^4}
Tr ( \ln D_0^{-1}D - D_0^{-1}D + 1)  + V_2(D),
\label{vfull}
\end{equation}
where  $D (D_0)$ is the complete(bare) gluon propagator, $V_2(D)$ is a two-particle
irreducible vacuum diagram and
the equation
\be
\frac{\d V}{\d D}=0,
\label{delv}
\ee
gives the SDE for gauge bosons in the Mandelstam's approximation.

The two-loop contribution to $V_2 (D)$ is given by
\be
V_2 (D) = \frac{-\imath}{6} Tr(\G^{(3)}D\G^{(3)}DD),
\label{ommand}
\ee
where $\G^{(3)}$ is the trilinear gauge boson coupling~\cite{mont}, and  in Eq.(\ref{ommand}) we
have not written the gauge and Lorentz indices, as well as the momentum integrals.

The vacuum energy density is given by the effective
potential calculated at minimum subtracted by its
perturbative part, which does not contribute to dynamical
mass generation ~\cite{cornja,cornor}
\be
\< \Omega \> = V_{min}(D) - V_{min}(D_p),
\label{omega}
\ee
where $D_p$ is the perturbative counterpart of $D$.  It is easy to see that \cite{gorbar}

\begin{equation}
\hspace{-0.5cm}\< \Omega \> = - \frac{3(N^2 -1)}{2} \int
\frac{d^4P}{(2\pi)^4} \, \left[ \frac{\P}{P^2+\P} - \ln \left(
1+\frac{\P}{P^2} \right) + \frac{2}{3} \frac{\P^2}{P^2(P^2+\P)} \right], 
\label{omeg}
\end{equation}
where all the quantities are in Euclidean space, $N=3$ for QCD and $\P$  is the gluon
 polarization tensor.
We will compute Eq.(\ref{omeg}) with $\P = {\mathcal M}^2 (P^2)$, where   $ {\mathcal M}^2 (P^2)$ is given by Eq.(\ref{ope})
with  $m^{\prime 2}= 0.50 \,\mbox{GeV}^2$, with the coupling constant, $\alpha(\mu^2=10 \, \mbox{GeV}^2)=0.20$. We obtain 
%%%%%%
%
\be
\frac{ \< \Omega \>}{\Lambda^4_{QCD}} = 3.60 \,.
\label{valomega}
\ee
with
$m^{\prime}/\Lambda_{QCD}= 3.90$.

Let us recall that the  vacuum expectation value of the trace of the energy
momentum tensor of QCD is~\cite{cre}
\be
\< \Theta_{\m\m} \> = \frac{\b(g)}{2g} \< G_{\m\nu} G^{\m\nu} \>,
\label{tmn}
\ee
where the perturbative $\b (g)$ function up to two loops is given by Eq.(\ref{beta}),
where the coefficients $\beta_0$ and $\beta_1$ are expressed in Eq.(\ref{coef}), 
and with $\alpha(\mu^2) = g^2 (\mu^2) / 4\pi$. 

We can relate the Eq.(\ref{tmn}) to the vacuum energy through
\be
\< \Omega \> = \frac{1}{4} \< \Theta_{\m\m} \> . 
\label{vt}
\ee

This  last expression for $\< \Omega \>$ can be compared
with the value of Eq.(\ref{valomega}) and in this way we obtain one estimative of the gluon
 condensate.

Using Eq.(\ref{beta}) and Eq.(\ref{coef}) with $n_f = 5$ (assuming that the inclusion 
of fermions
does not change drastically our results) and using $\alpha(\mu^2=10 \, \mbox{GeV}^2)=0.20$
 (equivalent to $m^{\prime}/\Lambda_{QCD}= 3.90$) we have found the following value
\begin{equation}
\gc =  0.015 \,\,\, \mbox{GeV}^4.    
\end{equation}

We obtained this last value  with $\Lambda_{QCD} = 182 \,\,\mbox{MeV}$ in the
left hand side of Eq.(\ref{valomega}) consistent  with the perturbative value shown in the third column of Table I. The outcome of this procedure is quite close to the value commonly used in QCD sum rules \cite{svz}: 
\begin{equation}
\gc =  0.012 \,\,\, \mbox{GeV}^4. 
\end{equation}

The above result indicates that our approximation gives a reliable estimative of the
 vacuum energy and that the inclusion of ghosts possibly do not modify the value of the vacuum energy \cite{gg}.
 
\section{Conclusion}

We computed the SDE for the gluon propagator in the Landau gauge within the Mandelstam
approximation where the fermions and ghost fields are neglected, and where the full gluon
vertex is inversely proportional to the gluon renormalization function.

The full triple gluon vertex is also extended to include the possibility of dynamical mass
generation, according to a prescription formulated by Cornwall many years ago. This 
prescription does not modify the SDE but is responsible for the compatibility of a massive
 gluon propagator
with the Slavnov-Taylor identity.

The renormalization of the SDE follows a procedure similar to the one proposed by Cornwall in
the renormalization of the SDE in the light cone gauge. It is particularly suited for a massive case
and leads to a renormalization constant of the form $Z_3 = 1 + f(\mu^2, \Lambda^2)$.

We were able to obtain a numerical solution in the full range of momenta that we have
considered without the need of introducing any asymptotic expression for the solution. The
propagator is renormalized using the central value of the coupling constant
at the b quark mass. The solution has been checked  to be stable within a  momentum 
range of twelve orders of magnitude. 

We verified that the ratio between the dynamical gluon mass and the QCD scale 
($m/\Lambda_{QCD}$) up to a large extent is independent on the choice of the renormalization point, and its value ($m/\Lambda_{QCD} \sim 2 - 3$) is consistent with previous estimates for this
 mass \cite{cornwall}.

Using a fit to the numerical solution we computed the vacuum energy and associated it with
the gluon condensate. The value that we obtain is consistent with the one usually assumed in
QCD sum rules.

Our calculation is far from being complete and the most obvious extension is the introduction of ghosts. An analysis of this case shows that the behavior of the gluon propagator is not
modified by the inclusion of the ghosts fields. However the behavior of the running coupling
constant as $q^2 \rightarrow 0$ may be different from zero as happens in the present case
\cite{gg}.
 
\section{Acknowledgments}

We benefited from discussions with A. Cucchieri and G. Krein and we would also like to 
thank A. Colato for his numerical hints. This research was supported by the Conselho Nacional 
de Desenvolvimento Cient\'{\i}fico e Tecnol\'{o}gico (CNPq) (AAN) and by
Funda\c{c}\~ao de Amparo \`{a} Pesquisa do Estado de S\~{a}o
Paulo (FAPESP) (ACA).

\end{document}